\documentclass[%
 aip,
 jap,%
 amsmath,amssymb,
 reprint,%
]{revtex4-1}
\usepackage{float}
\usepackage{graphicx}
\usepackage{dcolumn}
\usepackage{bm}

\usepackage[utf8]{inputenc}
\usepackage{gensymb}
\usepackage{xcolor}
\usepackage{ulem}
\usepackage{url}
\usepackage[colorinlistoftodos]{todonotes}
\usepackage{lineno}
\makeatother

\begin{document}


\title{Thickness dependence of magnetization reversal and magnetostriction in FeGa thin films}
\author{W. Jahjah}
\author{R. Manach}
\author{Y. Le Grand}
\author{A. Fessant}
 \affiliation{Laboratoire d'Optique et de Magn\'etisme (OPTIMAG), EA 938, IBSAM, Universit\'e de Bretagne Occidentale, 29200 Brest, France}

\author{B. Warot-Fonrose}
	\affiliation{Nanomaterials Group - CEMES CNRS-UPR 8011, Universit\'e de Toulouse, 31055 Toulouse, France}
\author{A.R.E. Prinsloo}
\author{C.J. Sheppard}
   \affiliation{Department of Physics, Cr research group, University of Johannesburg, PO Box 524, Auckland Park 2006, South Africa}
  \author{D.T. Dekadjevi}
    \affiliation{Laboratoire d'Optique et de Magn\'etisme (OPTIMAG), EA 938, IBSAM, Universit\'e de Bretagne Occidentale, 29200 Brest, France}
    \affiliation{Department of Physics, Cr research group, University of Johannesburg, PO Box 524, Auckland Park 2006, South Africa}
\author{D. Spenato}
\author{J.-Ph. Jay}
\email{jay@univ-brest.fr}

  \affiliation{Laboratoire d'Optique et de Magn\'etisme (OPTIMAG), EA 938, IBSAM, Universit\'e de Bretagne Occidentale, 29200 Brest, France}

\date{\today}

\begin{abstract}
Among the magnetostrictive alloys the one formed of iron and gallium (called "Galfenol" from its U.S. Office of Naval Research
discoverers in the late 90's) is attractive for its low hysteresis, good tensile stress, good machinability and its  rare-earth free composition. One of its applications is its association with a piezoelectric material to form a extrinsic multiferroic composite as an alternative to the rare room temperature intrinsic multiferroics such as  BiFeO$_3$.
This study focuses on thin  Fe$_{0.81}$Ga$_{0.19}$   films of thickness 5, 10, 20 and 60 nm deposited by sputtering onto glass substrates. Magnetization reversal study reveals a well-defined symmetry with two principal directions independent of the thickness. The magnetic signature of this magnetic anisotropy decreases with increasing FeGa thickness due to an increase of the non-preferential polycrystalline arrangement, as revealed by transmission electron microscopy (TEM) observations. Thus when magnetic field is applied along these specific directions, magnetization reversal is mainly coherent for the thinnest sample as seen from the transverse magnetization cycles.
Magnetostriction coefficient reaches 20 ppm for the 5 nm film and decreases for thicker samples, where polycrystalline part with non-preferential orientation  prevails.

\end{abstract}



\maketitle



\section{Introduction}

Magnetostrictive thin films have attracted a huge  interest in the last few decades. Indeed, it is of interest to understand the fundamental mechanisms driving the strain and magnetic coupling in thin films. Also,  magnetostrictive properties are the key properties used in different applied issues and devices such as sensors, actuators, energy-harvesters, spintronic devices\cite{mudivarthi_anisotropy_2010,jen_magneto-elastic_2014} and straintronic devices \cite{roy_hybrid_2011,ahmad_reversible_2015}.
For example a  new   technology,  called AAMR (acoustically
assisted  magnetic  recording) based on FeGa  has been recently proposed. This technology  uses  a  surface
acoustic wave (SAW) to modulate the coercivity of the recording  medium  by  the  inverse  magnetostrictive effect \cite{li_acoustically_2014}.

Among the different magnetostrictive alloys, FeGa discovered in 2000 by Clark \cite{clark_magnetostrictive_2000}   has received a particular attention as it exhibits remarkable properties such as low hysteresis, large magnetostriction, good tensile strength, machinability and recent progress in commercially viable methods of processing \cite{atulasimha_review_2011}.
Although FeGa magnetostrictive properties are lower than compared to those of Terfenol-D (a terbium-iron-dysprosium alloy), gallium, when substituted for iron  increases the tetragonal magnetostriction coefficient $\lambda_{100}$ over tenfold. This increase is generally attributed to the formation of Ga pairs when the Ga concentration reaches a substitution concentration value of 19\% Ga whereas it is commonly believed that  the formation of an ordered DO3 phase, as the alloy composition approaches 25\% Ga, is detrimental to magnetostriction \cite{lograsso_structural_2003}.

These FeGa alloys have the main advantage of being free of rare-earth elements and, thus, the cost is reduced compared to the rare-earth alloys family that has  another drawback which is brittleness.
The bulk FeGa phase diagram is rich since many phases can coexist. Among them, on the Fe rich side, one can cite : the disordered bcc A2 phase, the ordered cP B2 phase and  the ordered DO$_3$ phase \cite{okamoto_fe-ga_1990,ikeda_phase_2002}.

FeGa can be a good candidate to be associated with a piezoelectric material to form a composite multiferroic material. Such  "extrinsic multiferroic" materials are an alternative path to the intrinsic multiferroics since only a few of the latter have high enough critical temperatures to be used in devices\cite{parkes_magnetostrictive_2013, spaldin_NM_2019_Advances}.

In composite materials, properties of each layer are combined : for example magnetization can be controlled by an electric field through a strain transfered from the piezoelectric material to the magnetostrictive one. This is the converse magnetoelectric effect (CME) which can be schematically written as CME= (electric/mechanical) $\times$ (mechanical/magnetic).

For high frequency devices, in-plane uniaxial anisotropy is needed. One way to adjust the anisotropy has been recently adressed in FeGa films ($35$ to $55$~nm thick) deposited on PMN-PT piezoelectric substrates.  Anisotropy is modified using both oblique incidence deposition and electric field \cite{zhang_electric-regulated_2018}.

A magnetoelectric composite composed  of  a  layer  of  FeGa  associated with  a  layer  of  PZT-5H  have been  studied  for  potential   applications  in surgery such as cutting tools \cite{wang_thickness_2010}.

Clark \textit{et al}. \cite{clark_extraordinary_2003} have shown a  non-monotonic dependence of magnetostriction with Ga concentration with a double peak : the first one for 19~\% Ga content with $\frac{3}{2} \ \lambda_{100}$ reaching 400 ppm, the other peak is close to 27~\% with a slightly smaller magnetostriction coefficient. 
Theses authors have also shown a  sharp change from ${\frac{3}{2} \lambda_{111} \approx -20}$~ppm  to ${\frac{3}{2}\lambda_{111} \approx +40}$~ppm around 20~\% Ga alloy composition.

Very recently,  it has been found that  FeGa  single crystals with specific thermal treatment  behave in an unusual way: they show almost no magnetic hysteresis and these alloys are  "non-Joulian" and their overall volume is not conserved during magnetostriction process \cite{chopra_non-joulian_2015,james_materials_2015} contrary to the usual behavior when  the material distorts in shape but not in volume since while it expands in one direction it contracts in the transverse directions.  These alloys could find new applications in sensors and actuators.
The magnetostrictive properties of FeGa alloy depend strongly on composition preparation methods and thickness. 





Previous works on thin films (thickness of 150 nm) have shown the same double peaked maximum with Ga concentration as observed in bulk \cite{basantkumar_integration_2006} 
 but with lower magnetostriction values in 500 nm thick films \cite{hattrick-simpers_combinatorial_2008}.
 Javed \textit{et al}. have studied the effect of sputtering deposition conditions on 50 nm thick films with Ga concentration  ranging from 19 \% to 23\% and found a $<110>$ crystallographic texture normal to the film plane and an effective saturation magnetostriction close to 60 ppm \cite{javed_structure_2009}.
Javed \textit{et al}. also studied the thickness dependence (20--100 nm) of magnetostriction in a Fe$_{80}$Ga$_{20}$ polycrystalline alloy and determined surface magnetostriction contribution \cite{javed_thickness_2010}.

Epitaxially grown  Fe$_{81}$Ga$_{19}$  on Si/Cu with different thicknesses from 10 to 160 nm \cite{weston_fabrication_2002} or grown on Mgo(100) with 90 nm thickness films  have also been studied \cite{butera_growth_2005}.

This article intends to focus on magnetization reversal and magnetostriction of very thin films ($5 \leq$ thickness $\leq 60$ nm) prepared by sputtering deposition which to our knowledge has only been reported seldomly for thicknesses thinner than $10$~nm \cite{endo_effect_2017}.

\section{Experimental details}

In this paper, a study of ultrathin FeGa thin films grown by a sputtering technique is presented. FeGa/Ta  layers were  grown by standard RF diode sputtering onto glass substrates (Schott D 263 TM  \cite{schott_d263}). A Fe$_{81\%}$Ga$_{19\%}$ target was used  and the sample composition was checked with Electron Probe Micro Analysis (EPMA) measurements. The base pressure prior to the film deposition was typically 10$^{-7}$ mbar. The ferromagnetic (FM) FeGa thicknesses were $t_{FM} =  5,10, 20$ and $60$~nm. 
Ta is used  as capping layer to protect the FeGa from oxidation.
The RF power used to sputter the 3 inches FeGa target was $100$~W and the argon pressure was $1.5 \times  10^{-2}$ mBar. With these sputtering conditions the growth rate was 0.22 nm$\cdot$s$^{-1}$. This rate is close to the one found by Weston \textit{et al}. (0.3~nm$\cdot$s$^{-1}$) with quite similar deposition conditions \citep{weston_fabrication_2002}.

 An in-plane magnetic field of $H_\textsubscript{dep}\sim 2.4$ kA $\cdot$m$^{-1}$ ($\sim 300$~Oe) was applied, during deposition, to favor an uniaxial magnetic anisotropy. Structural analysis were performed by transmission  electron microscopy (TEM) experiments on cross-sectional lamellas, thinned by mechanical polishing and argon ion milling at low temperature using a Gatan Precision Ion Polishing System equipped with a liquid nitrogen cooling system. 
 Static magnetic measurements were performed  with a home-built
 vectorial vibration sample magnetometer (VVSM) allowing to determine both the longitudinal and transverse components of in plane magnetization at room temperature.
 
Coercive field temperature dependence was also obtained from $M(H)$ measurement using a Cryogenic cryogen free physical properties measurement platform with a vibrating sample magnetometer inset. The magnet was initially demagnetized after which the sample was cooled in zero applied magnetic field to the desired temperature. The $M(H)$ was then measured using the low magnetic field option of the Cryogenic system

Magnetostriction was characterized using optical deflectometry. Under an applied magnetic field, the cantilever (made of glass substrate and FeGa deposited  onto it) deformation was recorded through laser beam reflection at the free tip. Further details can be found in a previous paper \citep{jay_direct_2010}. In thin films, since the magnetostrictive sample is grown onto a much thicker substrate, the magnetostrictive deformations are hindered by this substrate and  one usually deals with magnetoelastic coupling coefficient, $b$ ,rather than magnetostriction coefficient, $\lambda$. This coefficient represents the characteristic magnetostrictive stress (see for example reference number \cite{jay_direct_2010}).

\section{XRD and TEM characterization}

\begin{figure}[h!]
\centering
\includegraphics[width=0.9\columnwidth]{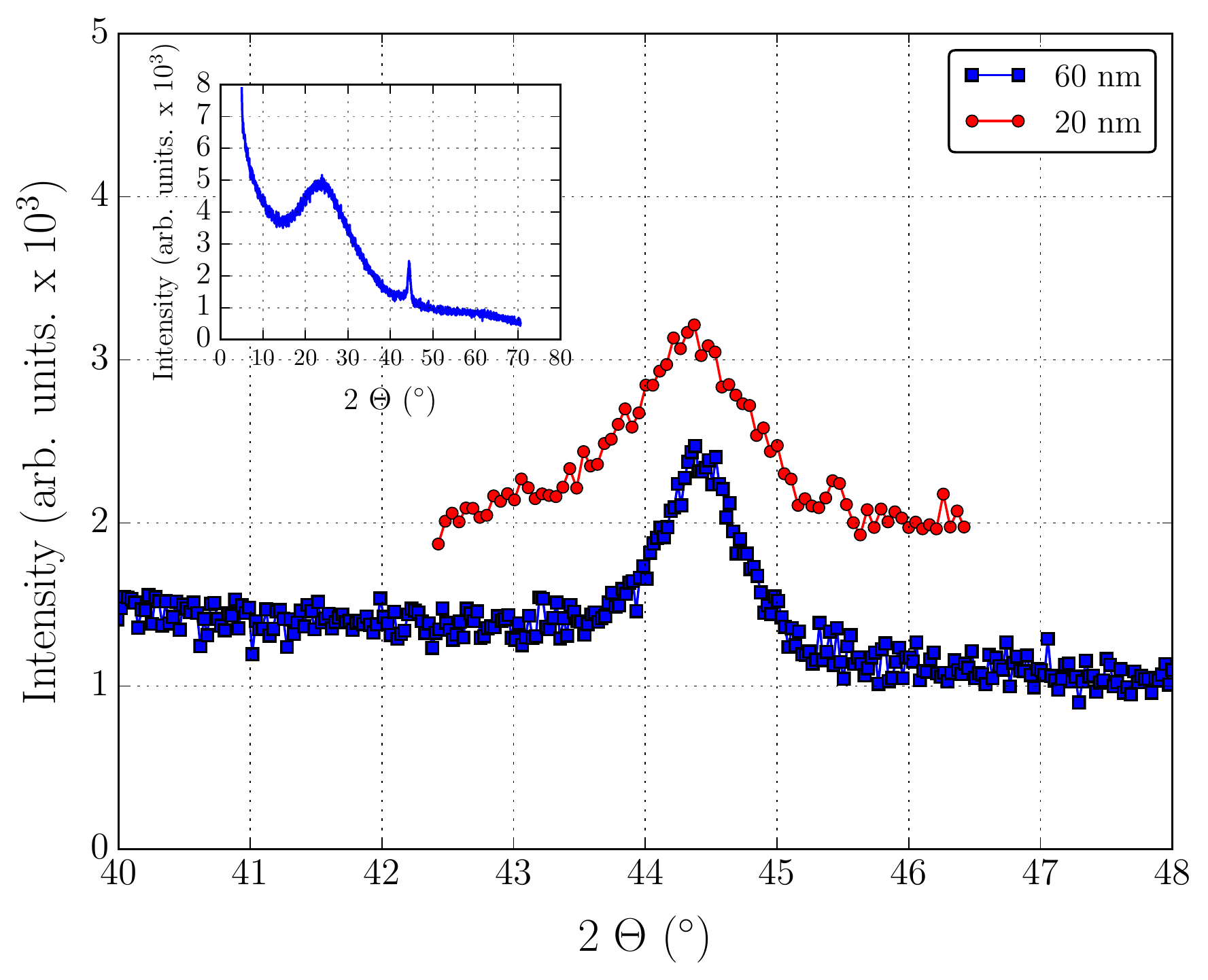}
\caption{\label{fig:XRD} $\theta-2\theta$ pattern for 60 nm and 20 nm thick FeGa films. Inset : full angle scan. }
\end{figure}

Crystallographic properties were first determined using X-ray diffraction as shown in figure \ref{fig:XRD}. A very broad hump 
is observed around $20^\circ$ which can be attributed to the X-ray signature of the amorphous glass substrate. A single Bragg peak is observed at 44.4 $^\circ$. This Bragg peak position corresponds to an inter-planar spacing of $0.288$~nm  which is in agreement with the results of Javed \textit{et al}. for a 50 nm film \cite{javed_structure_2009} who had noticed that the lattice parameter observed  in such thin films is lower than those of thicker films or bulk values.



This Bragg peak position does not change for FeGa thicknesses between 20 and 60 nm. It shows that the FeGa crystallographic lattice does not evolve with thickness. Below $20$~nm, it was not possible to measure clear XRD spectra because of the poor signal-to-noise ratio. 
It may be noted here that X-ray reflectivity was carried out on all samples to confirm the thicknesses. Nominal thicknesses appeared to be very close to those determined by reflectivity. These X-ray spectra exhibited well defined Kiessig fringes in agreement with low roughness thin films.

\begin{figure*}
\centering
\includegraphics[width=0.80\textwidth]{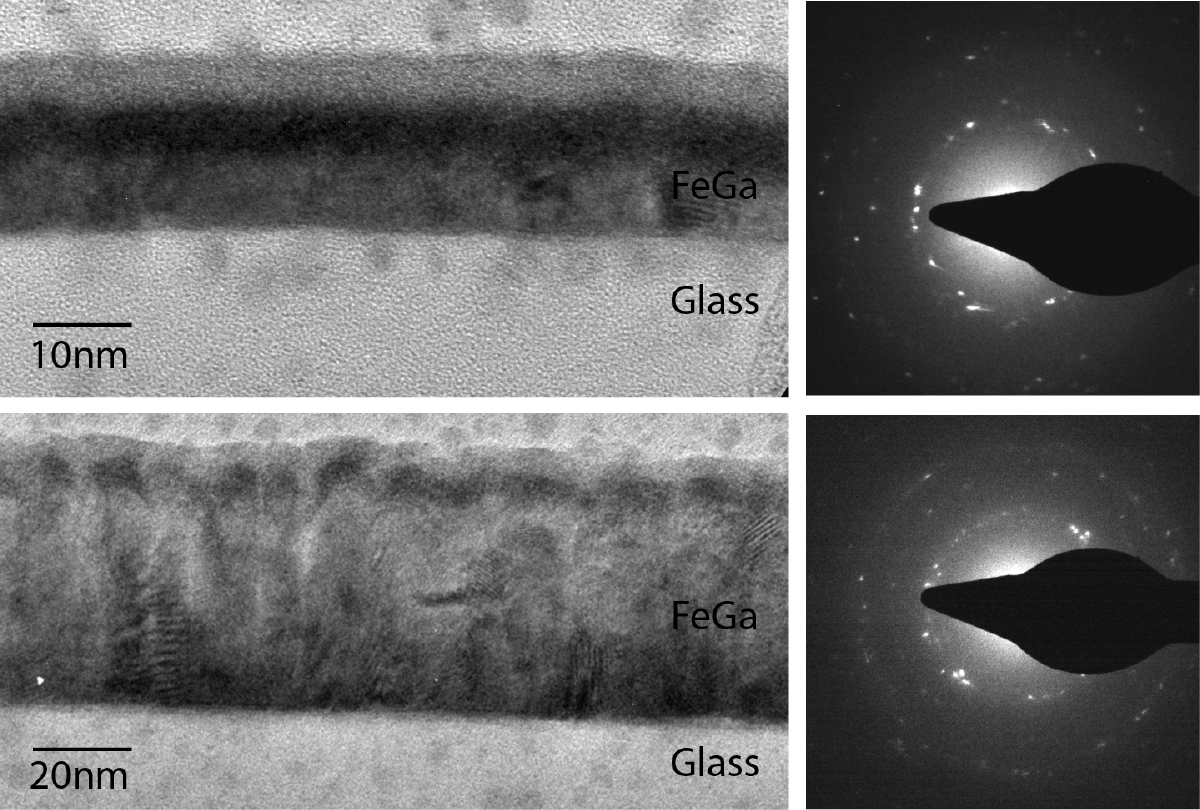}
\caption{\label{fig:TEM} High-resolution TEM image (left column: direct space, right column: reciprocal space) of a typical Glass/FeGa(7nm)/Ta bilayer is shown in the top image. The bottom figure shows the image obtained for the Glass/FeGa(40~nm)/Ta thick film. }
\end{figure*}

TEM characterization was carried  to probe the crystallographic properties for the thinnest films and also to determine the morphology of the films. TEM characterization is presented in figure \ref{fig:TEM} for the two samples of $7$ and $40$~nm thickness.

TEM images show that  the thin films are continuous with the absence of holes and  a low roughness for all thicknesses.
The  diffraction pattern exhibit non-uniform circular rings with well-defined points/peaks  present at some angular positions. Circular rings with a uniform distribution of intensity are a signature of a polycrystalline crystallographic systems with no preferential orientation. However, the presence of well defined spots in diffraction pattern is characteristic of a well-defined long range ordering encountered in epitaxial systems or substrates. The superposition of well defined spots and circular rings therefore indicates that the thin films are composed of a polycrystalline fraction with no preferential orientation and another part of the volume that grows in a precise direction as in epitaxial systems \cite{richy_JPDAP_2018_Temperature, cappuccio__1996_Thin}. 
The reinforcement of spot intensity in some directions is clear on the diffraction pattern corresponding to the thinnest sample, indicating a more "textured" film than the thicker one.

Such a fact may have critical consequences on the magnetic and magnetostrictive properties, which depend on the crystallographic arrangement, as will be shown later in this paper. In the following sections, the magnetic properties are analyzed using magnetometry and magnetostrictive techniques in the light of the structural characterization presented in this section.

\section{Magnetization reversal}
\begin{figure}[h!]
\centering\includegraphics[width=0.99\columnwidth]{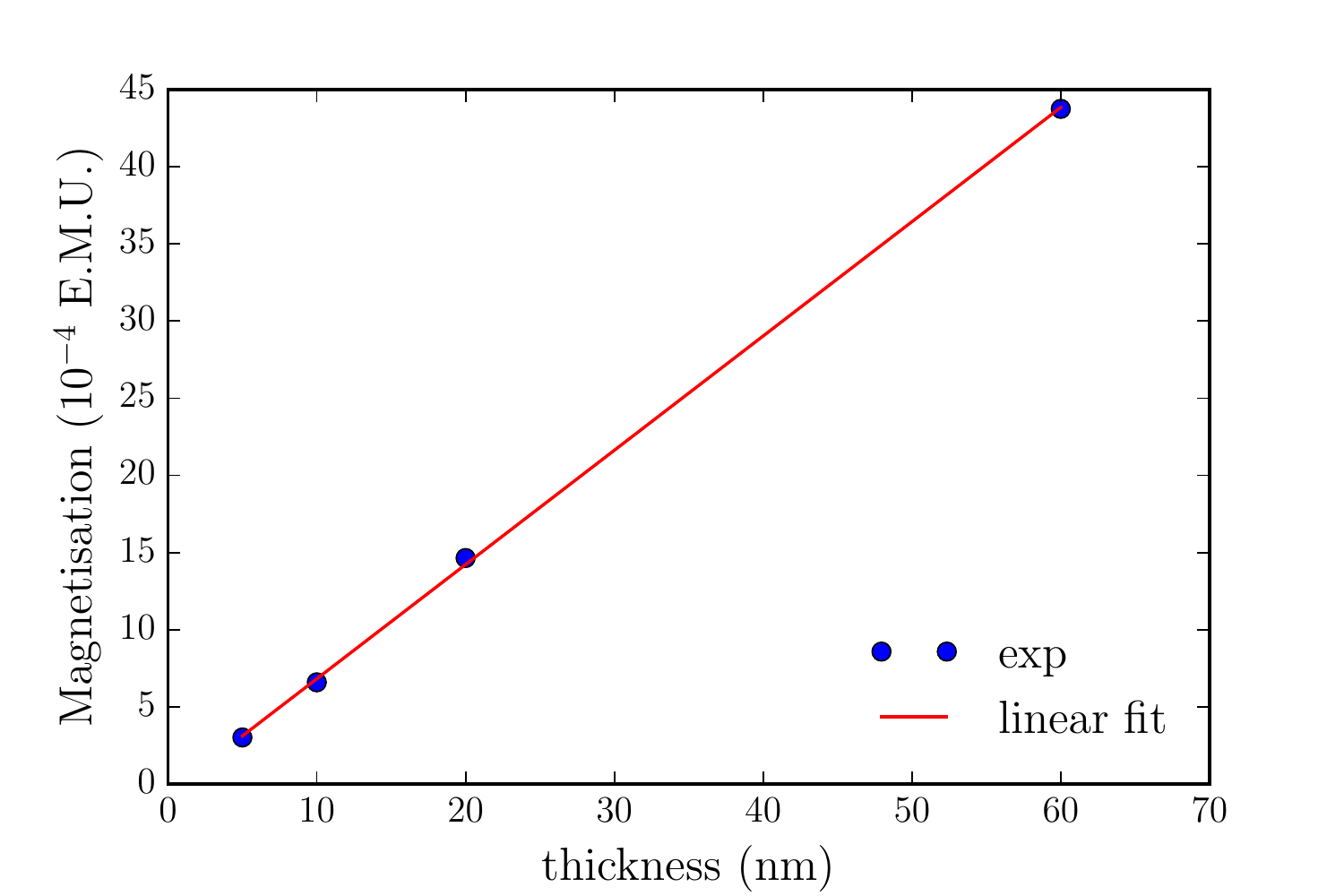}
\caption{\label{fig:M_thickness} Saturation magnetization  as a function of FeGa film thickness. Solid line is a linear fit.}
\end{figure}
\begin{figure}[h!]
\centering\includegraphics[width=0.99\columnwidth]{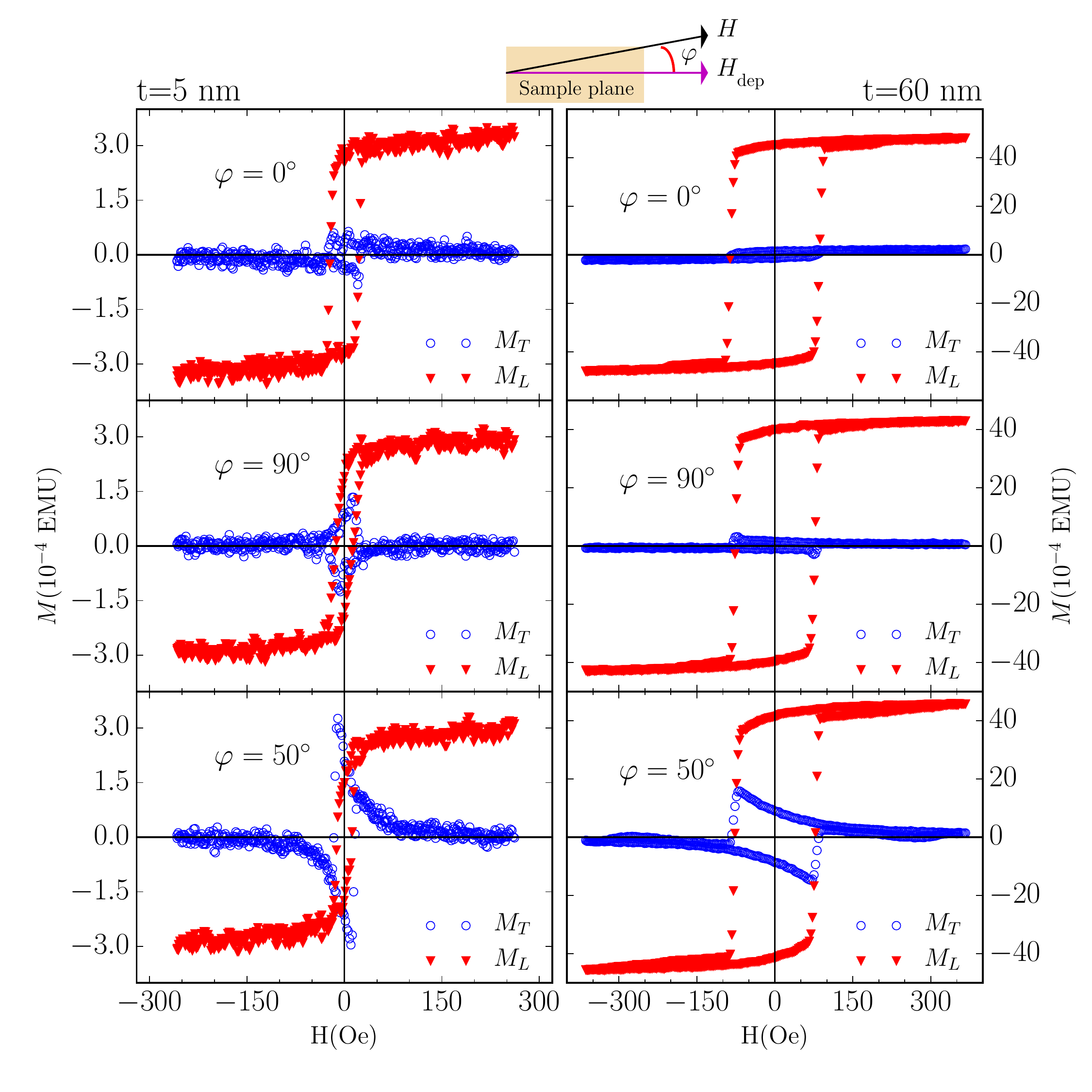}
\caption{\label{fig:M_H} In-plane magnetization loops of thickness 5~and~60~nm. Longitudinal $M_L$ (red~\textcolor{red}{$\blacktriangledown$}
) and transverse $M_T$  (blue~\textcolor{blue}{$\circ$}) components are shown. $\varphi$ is the angle between the deposition axis and the direction of the in plane  applied field. $\varphi=0$ corresponds to $H$ applied parallel to the direction of the deposition field.
}
\end{figure}

Magnetic measurements presented in figure \ref{fig:M_H} were performed at room temperature using our vectorial vibrating sample magnetometer.  Both longitudinal magnetization ($M_L$), the component parallel to the in-plane applied magnetic field $H$ and transverse magnetization ($M_T$), the component in the film plane, but perpendicular to $H$ have been measured with four detection coils. Figure \ref{fig:M_thickness} shows the evolution of the saturation magnetic moment as a function of the FeGa thickness. All the magnetic measurements were performed on samples with the same substrate surface area. The linear evolution of the saturation magnetic moment,  thus, means that all samples from 5 nm to 60 nm have the same saturation magnetization $M_s= 919$ kA $\cdot$ m$^{-1}$ ($\mu_0 M_s = 1.15$ T). This value is close to the one found by Zhang \cite{zhang_effect_2013} for a $50$~nm FeGa thick sample,  but smaller than the one found by Gopman in thin films of thicknesses ranging from 20 to 80 nm \cite{gopman_static_2017} or in bulk materials that showed a saturation magnetization of $1.5$ T. 

Even though Endo \textit{et al}. \cite{endo_effect_2017} have also observed a quasi-constant magnetization for Fe$_{0.78}$Ga$_{0.22}$ films with thicknesses between 3 and 100 nm, it is quite remarkable that the magnetization for the present sample series remains constant for all thicknesses since  a reduction of magnetization  at nanoscale may be observed  when the thickness is decreased i.e. reduction of the magnetization should be related to low dimensionality effects \cite{le_graet_driving_2016}.

\begin{figure}[h!]
\centering
\includegraphics[width=0.99\columnwidth]{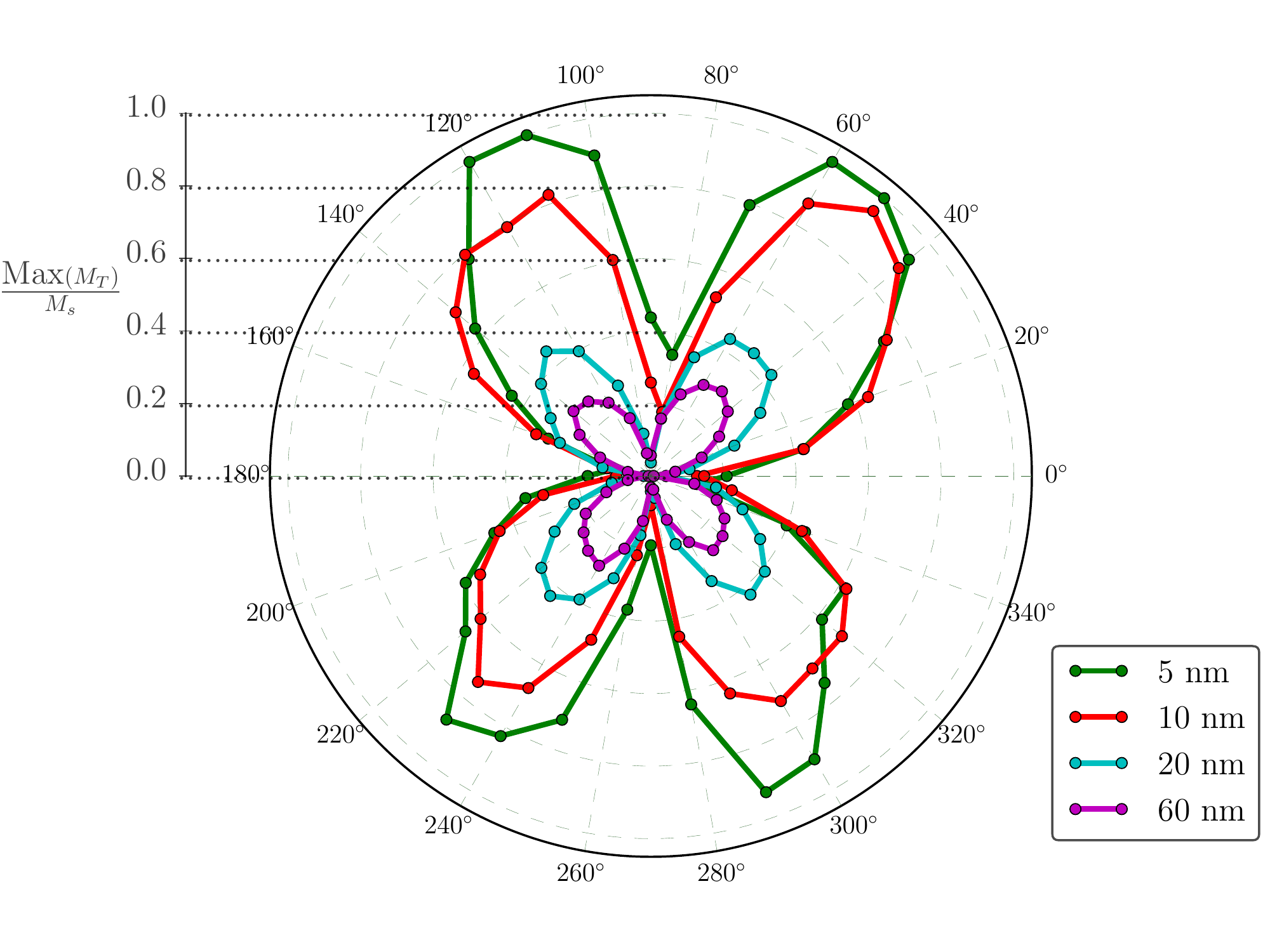}
\caption{\label{fig:Mt} Azimuthal behavior of the normalized maximum transverse magnetization, $\frac{\mbox{Max}(M_T)}{M_s}$, for different FeGa film thicknesses.  }
\end{figure}

Figure \ref{fig:M_H}  presents the magnetization hysteresis loops for the $5$~nm and $60$~nm samples for both transverse and longitudinal components. In-plane magnetic field is applied along the direction of the applied deposition field ($\varphi=0 ^\circ$) or perpendicularly to it ($\varphi=90 ^\circ$). The third angle ($\varphi=50 ^\circ$) corresponds to the maximum of $\mbox{Max}(M_T)/M_s$ found when varying $\varphi$ (see figure \ref{fig:Mt}).

\begin{figure}[h!]
\centering
\includegraphics[width=0.99\columnwidth]{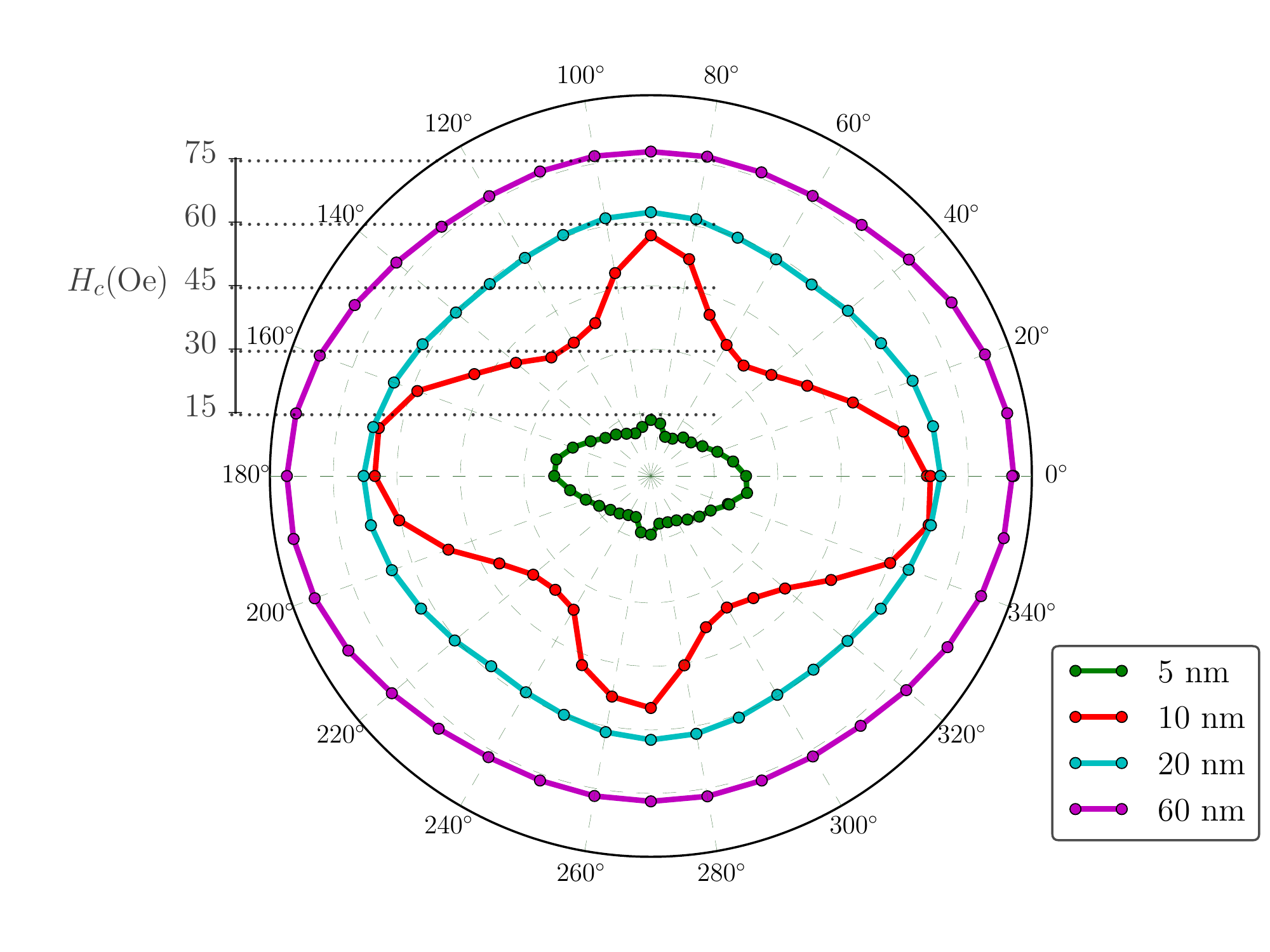}
\caption{\label{fig:HC} Azimuthal behavior of the coercive field $H_c$ for different FeGa film thicknesses.}
\end{figure}

For the $5$~nm sample and along the $\varphi=0$ direction, the transverse magnetization is flat showing the absence of a net  moment in the direction perpendicular to the applied field.  Thus,  the magnetization reversal is incoherent, involving domains nucleation and propagation. On the other hand, when external field is applied along the  $\varphi=50^\circ$ direction, the $5$~nm normalized transverse magnetization shows a maximum value of one, indicating a coherent magnetization reversal (i.e. a coherent (uniform) magnetization reversal)\cite{sellmyer_HTF_2002_Chapter}.  For larger thicknesses, the angular variation of transverse magnetization displays a similar behavior, in the sense of a maximum value at $\varphi=50^\circ$. However, this $M_T$ maximum value is reduced for  thicker samples. It should be noted here that the observed angular dependence of longitudinal and transverse component of magnetization does not correspond to an uniaxial anisotropic system with an easy axis induced by the applied field during growth. Indeed, such an uniaxial system would exhibit a maximum value of $M_t$ perpendicular to the magnetic field applied during growth. In order to understand the driving mechanism for the magnetic properties angular dependence, systematic in-plane azimuthal measurements were performed for all samples in increment of  $\varphi=10^\circ$. For each applied magnetic field direction, the maximum value of the transverse magnetization and the coercive field were obtained, as shown in figure \ref{fig:Mt}  and figure \ref{fig:HC}.

Figure \ref{fig:Mt}  shows the azimuthal angular dependence of the normalized (to the saturation magnetization $M_s$) maximum transverse magnetization $\mbox{Max}(M_{T})/M_{s}$ for all samples. It reveals a "X" shape with 4 lobes and  a well-defined symmetry: for a given thickness the 4 lobes have the same size and their maxima show up at about $50^\circ$, $120^\circ$, $230^\circ$ and $300^\circ$ with respect to the depositing field direction (i.e. $0^\circ$).
Even though the lobe amplitude decreases when thickness increases, it is worthwhile noting that their  orientation do not vary with thickness. The symmetry of the azimuthal angular dependence for the transverse magnetization relates to the magnetic anisotropy present in a system. An uniaxial anisotropy system with coherent magnetization rotation, as described by the Stoner-Wohlfarth model \cite{stoner_mechanism_1948} will exhibit a well-known azimuthal shape of $M_T$, with two maxima. However, a magnetic system with a predominant cubic anisotropy will exhibit an azimuthal shape with four maxima such as observed here. Furthermore, if the cubic anisotropy is of crystalline origin, then the maxima are related to the in-plane crystalline directions. Here, the angular positions of the maxima observed in figure \ref{fig:Mt} correspond to the angles between $[111]$ directions within the $(110)$ plane in the cubic system.


Considering that the preferential growth discussed previously is along the $[110]$ directions, it shows that the crystalline ordering is the driving mechanism for ultrathin sputtered FeGa systems studied here. FeGa system exhibit magnetically hard axis along defined crystallographic directions. It favors a coherent rotation mechanism (i.e. a maximum of the transverse component) along these directions as to minimize the anisotropic energy of the system. As already noted, the maxima orientation is not thickness dependent, whereas the  maxima intensity decreases strongly with increasing the thickness. It can be explained by an increase of the volume of the no-preferential polycrystalline arrangement with increasing the FeGa thickness, as revealed by TEM observations. Therefore, this  clearly demonstrates the presence of a surface anisotropy due to low dimensional effect in the RF sputtering ultrathin films. This surface anisotropy originates from thickness dependent crystallographic texture effect. 
The same kind of anisotropy was observed by Weston \textit{et al}.\cite{weston_fabrication_2002}  in epitaxially grown FeGa films with thicknesses ranging from 20 to 160 nm but not in sputtered ones : for very thin layer Endo \textit{et al}. have shown a small uniaxial anisotropy if $t \leq$~7.5 nm, but isotropic behavior above this thickness \cite{endo_effect_2017}.

It is of interest to determine the angular dependence of the coercive field as to further understand the effect of the magnetocrystalline component. The reason for this is such a property  is of prime interest  for FeGa based MEMS when studied in an external magnetic field \cite{basantkumar_integration_2006}. 

Figure \ref{fig:HC} shows a clear thickness dependence of the coercive field azimuthal shape. The thicker films exhibit a quasi-circular (isotropic) shape, whereas the thinnest ones show a more complex behaviour  with local extrema position reversed as compared to the $M_T$ evolution. Thus, the directions of the $H_c$ maxima correspond to $M_T$ minima and vice-versa.
When increasing thickness the maximum of $H_c$ increases from $20$ Oe to $75$ Oe.
These values are of the same order than those found in FeGa thin films grown epitaxially on Cu on Si \cite{weston_fabrication_2002} but at least twice smaller than the values found by Javed \cite{javed_structure_2009} in $50$~nm thick samples grown by co-sputtering and evaporation techniques.

Zhang \textit{et al}. \cite{zhang_effect_2013} deposited a $50$~nm thick FeGa onto PET flexible substrate with varying Ta buffer layer thicknesses. They observe an uniaxial anisotropy decreasing with Ta thickness, attributed to residual stress from deformations in PET substrate. Along the hard axis, the coercive field is reduced to 14 Oe when the Ta buffer layer is $20$~nm.
More recently the same group  studied wrinkled FeGa films ($20,40, 60$~nm) fabricated on elastic polydimethylsiloxane (PDMS) using two fabrication methods \cite{zhang_surface_2016}. They found more or less uniaxial anisotropy depending on the fabrication methods and coercive field ranging from $40$ to $90$~Oe.

Finally, it should be noted that the minima of the coercive fields may be used  in future to minimize the external energy needed for controlling devices.

In order to complement this room temperature magnetic study, coercive fields dependence with temperature from $10$~K to $300$~K was probed. This thermal behavior should be considered to understand the working temperature of sensor or actuator devices involving ultrathin FeGa films.

In order to measure the magnetization ($M$) of the FeGa films as function of applied magnetic field ($H$) at constant temperature the following measurement protocol was used. The protocol starts by a demagnetization of the super-conducting magnet, in order to obtain a zero magnetic field for the zero field cooled (ZFC) measurements. The thin film/glass are then cooled in zero field  to the desired temperature and after allowing the temperature to stabilize, $M(H)$ is then measured. After the $M(H)$ measurement, the system is heated back to $300$~K, and then ZFC cooled again to the next desired temperature. This protocol was repeated in $25$~K steps.

\begin{figure}[h!]
\centering
\includegraphics[width=0.99\columnwidth]{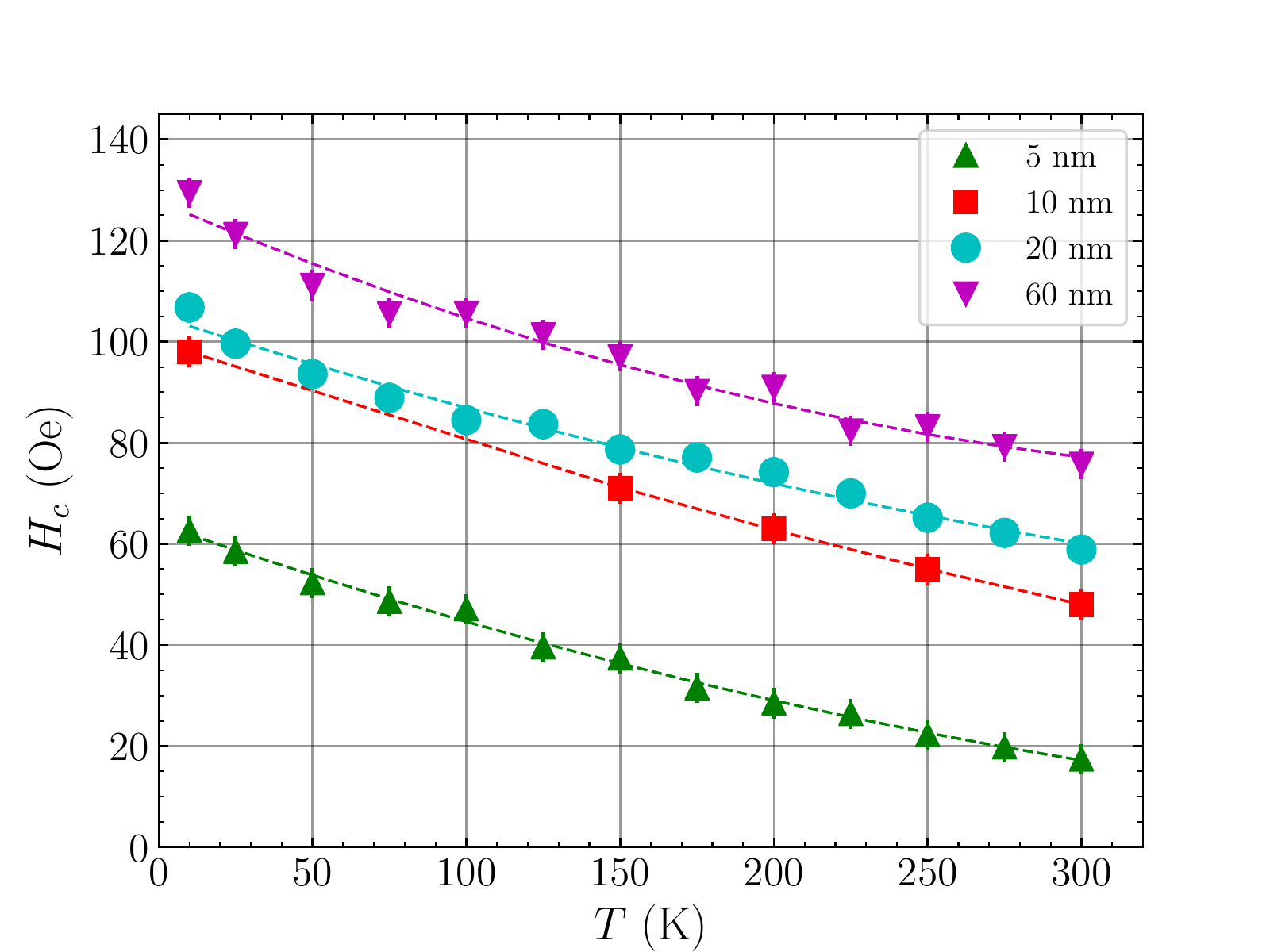}
\caption{\label{fig:Hc(T)} Temperature evolution of the coercive field, $H_C$,  for different FeGa film thicknesses.}
\end{figure}

Coercive fields, $H_c$, as a function of temperature for all the samples are presented in figure \ref{fig:Hc(T)}. The usual decreasing $H_c$ behavior with $T$ is observed as also reported in CoCrTa films \cite{zeng_studies_2002}, in exchange coupled NiFe/NiO layers \cite{dekadjevi_driving_2013} and in nanoparticles \cite{nunes_temperature_2004}.  

Coercive field as a function of temperature is  independent of the thickness and the relative ratio of $H_c$ for two different thicknesses  is conserved such that the thinnest sample  has always the smaller coercive field, while the thicker one remains the most coercive one for all temperature range. It indicates that there are no thickness dependent magnetic phase transition present in these FeGa films.

\section{Magnetostriction}

In a cubic crystal, the relative deformation due to magnetostriction is given by \cite{cullity_introduction_2009,petculescu_magnetoelasticity_2012} :

\begin{multline}
\label{delta_ell-1}
\frac{\delta \ell }{\ell}= \frac{3}{2} \lambda_{100} \left[ \alpha_1^2 \beta_1^2+ \alpha_2^2 \beta_2^2 + \alpha_3^2 \beta_3^2 - \frac{1}{3} \right] + \\
 3\, \lambda_{111}\left[ 
 \alpha_1 \beta_1 \alpha_2 \beta_2 + 
 \alpha_2 \beta_2 \alpha_3 \beta_3 + 
 \alpha_1 \beta_1 \alpha_3 \beta_3\right] 
\end{multline}

where $\lambda_{100}$ and $\lambda_{111}$ are  the saturation magnetostriction coefficients when the crystal is magnetized, and the strain is measured, in the directions $<100>$ and $<111>$, respectively \citep{cullity_introduction_2009}.
$\alpha_i$ and $\beta_i$ are respectively the direction cosines of the magnetization and observed elongation measured from cubic crystallographic directions ($[100],\, [010], \,[001])$.
The magnetostrictive strain can also be expressed in terms of the tetragonal ${\lambda^{\gamma,2}= \frac{3}{2} \lambda_{100}}$ and rhombohedral ${\lambda^{\epsilon,2}= \frac{3}{2} \lambda_{111}}$ magnetostriction constants, since an initially cubic cell magnetized along $[100]$ becomes slightly tetragonal, or rhombohedral when magnetized along $[111]$.

From equation (\ref{delta_ell-1}) it is clear that $\frac{\delta \ell}{\ell}$ is sensitive to crystallinity and therefore will be different for monocrystalline, polycrystalline or textured materials.

Let's call $x$ the sample thickness, $y$ the width and $z$ the length along which deformation is measured. When  magnetic field is rotated within the $y-z$ plane, one can compute the azimuthal variation of deformation assuming that for strong enough magnetic fields, magnetization remains parallel to it.
If $\varphi$  is the angle measured from the sample length  ($\varphi=0$ when $\vec{H} \parallel  z$ and $\varphi=90^\circ$ when $\vec{H} \parallel y$) one gets the general formula for the (averaged where appropriate) relative deformation: 
\begin{equation}
\frac{\delta \ell }{\ell} = A + B \cos(2 \varphi), \label{delta_ell-2}
\end{equation}

where 
$A$ and $B$ are constants including both $\lambda_{100}$ and $\lambda_{111}$ with weighting coefficients depending on the crystallinity of the sample, as we show below for three simple cases. 

In the case of a fully randomly oriented polycrystalline material, using Birss calculations	\cite{birss_saturation_1960} one finds: 
 ${A=\frac{1}{10}\lambda_{100}+ \frac{3}{20}\lambda_{111}}$ and ${B=\frac{3}{10}\lambda_{100}+ \frac{9}{20}\lambda_{111}}$.
 
 In the case of a single-crystal oriented such as  $[110]\parallel~x$, $[001]\parallel~y$ and $[\bar{1} 1 0] \parallel~z$ one finds:
 ${A=-\frac{1}{8}\lambda_{100}+ \frac{3}{8}\lambda_{111}}$ and ${B=\frac{3}{8}\lambda_{100}+ \frac{3}{8}\lambda_{111}}$.

In the case of a textured material with $[110] \parallel x$ and  random orientations within the $(y-z)$ film plane,  one finds:  ${A=\frac{1}{16}\lambda_{100}+ \frac{3}{16}\lambda_{111}}$ and ${B=\frac{9}{32}\lambda_{100} + \frac{15}{32}\lambda_{111}}$.


In thin films, magnetostriction deformations are hindered by the much thicker substrate. In our experiment we measure the bending angle $\theta(H)$ of  the cantilever tip when magnetic field is cycled from $H_{max}$ to $-H_{max}$ and back to $H_{max}$ with $H_{max} > H_{sat} $ where $H_{sat}$ is the saturation field when magnetization reaches its saturation value \cite{jay_direct_2010}.  
The bending angles $\theta(H)$ are converted to stress using the well-known formula \cite{lacheisserie_magnetostriction_1994} : 
\begin{equation}
b(H)=\frac{E_s t_s^2}{6(1+\nu_s) L \ t_f} \times \theta(H),
\end{equation}
\\where $E_s= 72.9$~GPa is the Schott D263 glass substrate Young modulus, and $\nu_s=0.208$ its Poisson ratio \cite{schott_d263}. 

The glass substrate thickness is $t_s=30~\mu$m and the FeGa film thickness is $t_f$. $L$ is the sample length (more precisely the distance between the support and the laser spot on the sample).
One can access the magnetostriction coefficient through $\lambda= -\frac{2}{3} \left( \frac{1+\nu_f}{E_f}\right) \times b$, but the elastic coefficients of the thin film ($\nu_f$ and $E_f$) are needed. 
For very thin films, these elastic parameters are difficult to measure and usually stress $b$ is considered as the relevant parameter for magnetostrictive effects. Nevertheless, following Hattrick-Simpers \cite{hattrick-simpers_combinatorial_2008},  we take $\frac{E_{f}}{1+\nu_{f}}=50$~GPa for FeGa to compute $\lambda$ and compare the values we get  with that obtained by other authors.

\begin{figure}[h!]
\centering
\includegraphics[width=0.99\columnwidth]{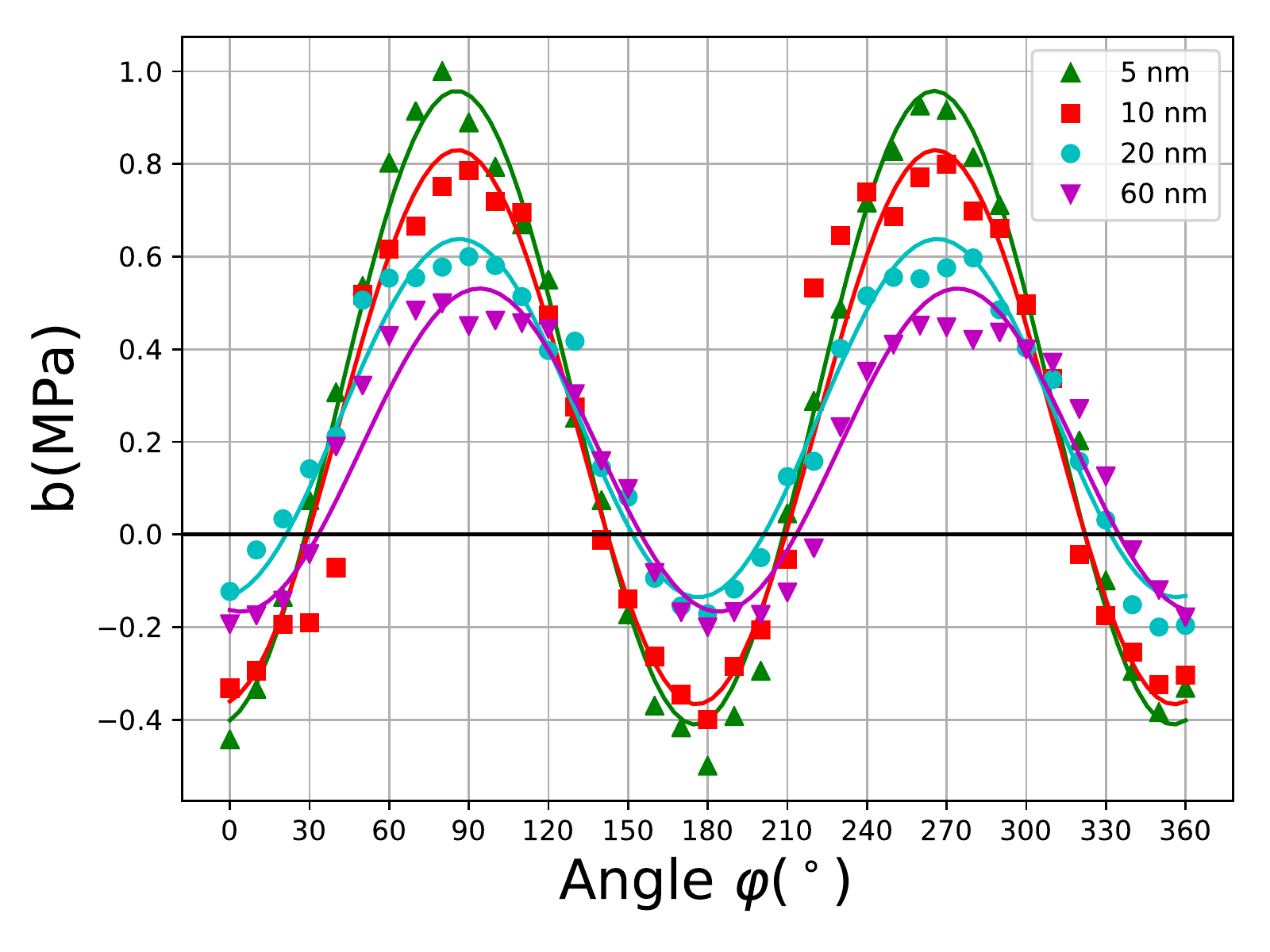}
\caption{\label{fig:Mst_azim} Azimuthal behavior of magnetostriction for {${t=5,10,20,60}$} nm thick films. For each thickness solid line is a fit to experimental points using the function ${A + B \cos(2 \varphi+ \delta)}$. 
}  
\end{figure}

The azimuthal behavior of $b$ is shown in figure  \ref{fig:Mst_azim}. The $\cos(2 \varphi)$ (where $\varphi$ is the angle between the applied magnetic field $H$ and the cantilever length) expected behavior (see equation (\ref{delta_ell-2})) is found for all sample thicknesses. 

\begin{figure}[h!]
\centering
\includegraphics[width=0.99\columnwidth]{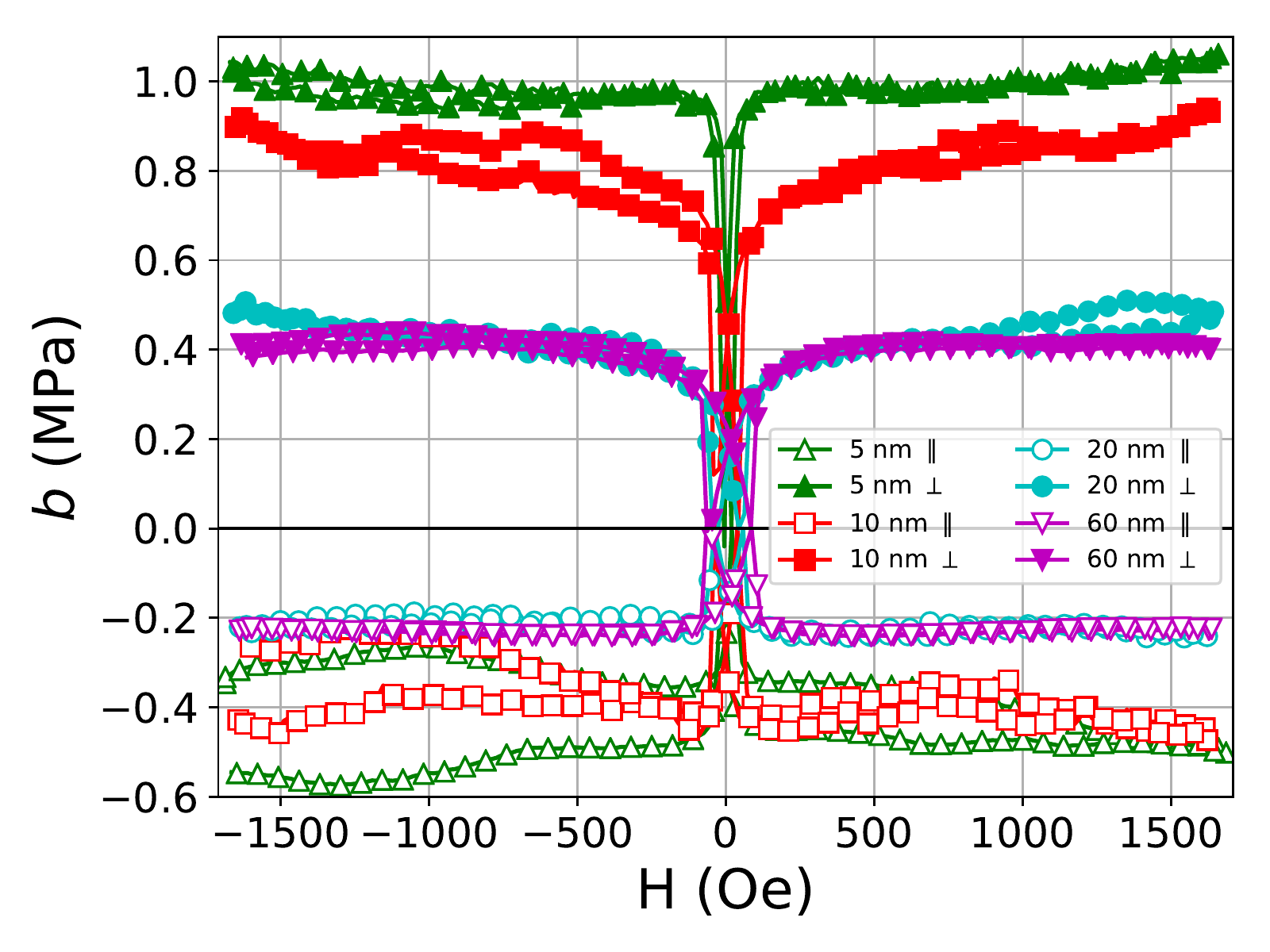}
\caption{\label{fig:bending} Bending $H$-cycles of FeGa cantilever films for thicknesses ${t=5,10, 20, 60}$~nm. Magnetic field is applied  either parallel to the cantilever length ($\parallel$) or perpendicular to it~($\perp$).}
\end{figure}

The full $H$-bending cycles for the 2 remarkable angles  ${\varphi=0}$ ($\parallel$ cantilever length and deposition field) and ${\varphi=90^\circ}$ ($\perp$)
are shown in figure  \ref{fig:bending}.

To access the effective magnetostriction constant ($\lambda_\textsubscript{eff}$) or the characteristic magnetostrictive stress ($b\textsubscript{eff}$) 
one has to perform at least two measurements: along the sample length (parallel) and along the sample width (perpendicular), since 
magnetostriction constant is given by \cite{cullity_introduction_2009,kaneko_MeasuringMagnetostrictionThin_1988}
$\lambda_\textsubscript{eff}=~\frac{2}{3}\left[ (\frac{\delta \ell}{\ell})_{\parallel} - (\frac{\delta \ell}{\ell})_{\perp}\right]$
and similarly ${b_\textsubscript{eff}=b_\parallel-b_\perp}$.
Obviously magnetostriction can also be deduced from the azimuthal behavior of $b$, i.e. the $\cos(2 \varphi)$ amplitude as seen in figure \ref{fig:Mst_azim}.

One should notice that $b_\textsubscript{eff}$ is negative for these FeGa samples. This corresponds to a positive magnetostriction since the magnetostriction coefficient $\lambda$  and  $b$ are related through a negative coefficient and thus a dilation along the applied field direction.

%
%
%
%

Figure \ref{fig:bending} shows the bending $H$-cycles for the different sample thicknesses. 
The   coercive field enhancement already seen with increasing thickness (measured from  $M-H$ VSM loops and shown in figure \ref{fig:Hc(T)}) is also visible on the  magnetostriction cycles. Indeed the magnetostrictive loops widen when thickness increases and the coercive field corresponds to the field when  $b$ switches to $0$.

\begin{figure}[h!]
\centering
\includegraphics[width=0.99\columnwidth]{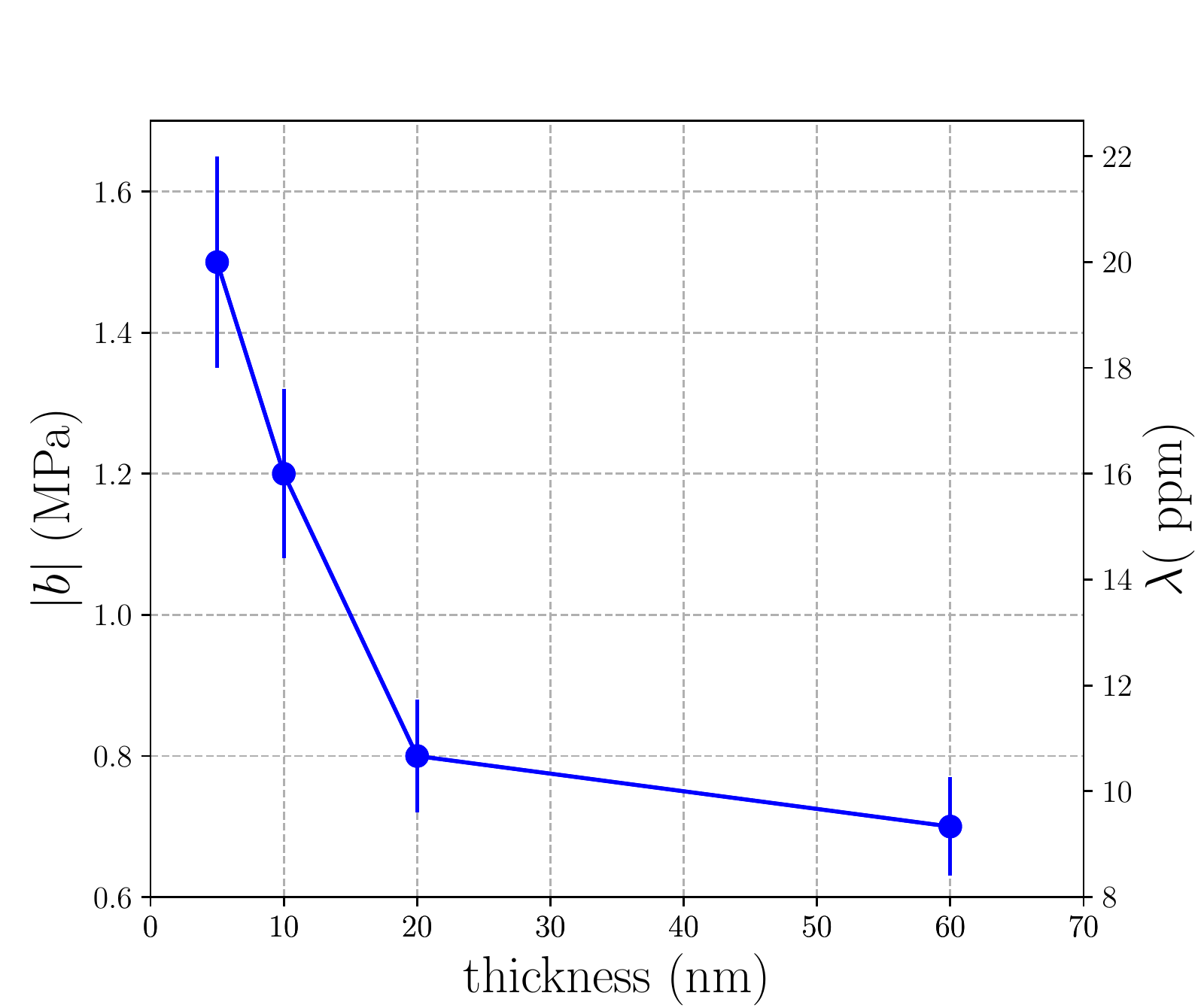}
\caption{\label{fig:bgamma(t)} Evolution of the magnetoelastic coefficient, $|b|$, and magnetostriction coefficient, $\lambda$, with FeGa thickness. The solid line is guide for the eye.}
\end{figure}

The evolution of $|b_{eff}|$ and $\lambda_{eff}$ with film thickness is presented in figure \ref{fig:bgamma(t)} which shows that $|b|$ is thickness dependent, since $|b_{eff}|$ (resp. $\lambda$) decreases from $1.5$~MPa to $0.7$~MPa (resp. from $20$~ppm to $9$~ppm) when $t$ increases from $5$~nm to $60$~nm. 

When decreasing thickness, our structural and magnetic studies have shown that sample texture increases. This structural modification impacts $\lambda_{eff}$ since the weighting coefficients of $\lambda_{100}$ and $\lambda_{111}$ appearing in $A$ and $B$ in equation (\ref{delta_ell-2}),  but it is not possible to exactly determine these coefficients since precise crystallography and orientations are not accurately known.
However the difference between the thinnest sample ($5$~nm) and the thicker samples cannot exclusively be explained by this effect.

In fact, for very thin samples, the proportion of interface/surface atoms with obviously different symmetry from bulk atoms is large. Thus this may contribute differently to magnetostriction, as observed in anisotropy where a surface term is usually added to the bulk one when the film in only constituted of a few  atomic planes. Similarly to anisotropy phenomenological expression,
one can write:
 $\lambda_\textsubscript{eff}=\lambda_\textsubscript{bulk}+\frac{\lambda_\textsubscript{surf}}{t}$, as suggested by N\'eel \cite{neel_anisotropie_1954} and Szymczak \cite{szymczak_surface_1993}.
Once again, the thickness evolution obtained in the present study  cannot be explained alone by only considering the surface contribution. It would be hazardous to extract $\lambda_\textsubscript{bulk}$ and $\lambda_\textsubscript{surf}$ from the present data since varying the thickness modifies the crystallinity.
The evolution of magnetostriction with thickness is probably a combination of both crystallinity and surface effects.

The order of magnitude  obtained in the present investigation  for magnetostriction is comparable with the value found by other authors for comparable Ga alloy concentrations (20~\%),  but the thickness evolutions are rather different according to various studies:
Javed \textit{et al}. \cite{javed_thickness_2010} found an increase of $\lambda\textsubscript{eff}$ from $\approx 20$ to 80 ppm when $t$ increases from 20 to 100 nm in Fe$_{80}$Ga$_{20}$ .
Yu \textit{et al}.\cite{yu_effect_2015} showed that magnetostriction decreases from 35 ppm to 28 ppm when thickness increases from 10 to 120 nm .

For very Fe$_{78}$Ga$_{22}$ thin samples (3 -- 10~nm) Endo \textit{et al}.\cite{endo_effect_2017} reported a magnetostriction fluction  between 15 and 20~ppm,  reaching  24~ppm for 30~nm and then decreasing to 18~ppm at 100~nm.

For a Fe$_{81}$Ga$_{19}$ $110$ nm  thick film deposited on glass, Jen \textit{et al}. \cite{jen_magneto-elastic_2013} found $\lambda_s=21$ ppm , a very close value to that measured by Yu \textit{et al}. in a sample deposited onto a flexible PET substrate \cite{yu_static_2015}. Thus, the value obtained in the present study for the magnetostriction corresponds well with that previously reported in literature.



\section{Conclusion}

We have shown that thin Fe$_{81}$Ga$_{19}$ films deposited by sputtering exhibit peculiar anisotropy directions even if growth is initiated onto amorphous glass substrate. This anisotropy revealed by XRD and TEM is prominent for the thinnest films ($t \leq 10$ nm) and have clear consequences on magnetic and magnetostrictive behaviors: when rotating the applied magnetic field in the film plane, transverse magnetization is maximum for specific directions  which in-between angles corresponds to the $[111]$ family directions angles in the $(110)$ plane. Thus, when external field is applied along these specific directions,  magnetization reversal is mainly coherent.  The present study also shows that magnetostriction coefficient is higher for the thinnest samples (5 nm), where the polycrystalline part of the sample is reduced as compared to the thicker one (60 nm). 
Film thickness thus strongly affects magnetic and magnetostritive properties of FeGa films. It has to be taken into account for applications when anisotropy is required, such as in microwave devices or when this magnetostrictive alloy is associated with a piezoelectric material as to form an extrinsic multiferroic composite where electric field is used to control magnetic properties.

\section{Acknowledgments}

R\'egion Bretagne is acknowledged for ARED support (partial W.J. PhD financial support).\\
This work was supported by South African National Research Foundation Grant No’s: 80928, 93551, 115346, 80880) and the URC and FRC of the University of Johannesburg, South Africa. 


\section{Bibliography}
\bibliography{ref}

\end{document}